\documentstyle[mnras_cite,epsfig,eqnarray]{mn2e}
\title[Spin Properties of Supermassive Black Holes]
{Spin Properties of Supermassive Black Holes with Powerful Outflows} 
\author[R. A. Daly]{Ruth. A. Daly\thanks{E-mail: rdaly@psu.edu}\\
Penn State University, Berks Campus, Reading, PA 19608, USA}
\begin{document}



\maketitle

\label{firstpage}

\begin{abstract}
Relationships between beam power and accretion disk luminosity are studied for a sample of 55 HERG, 13 LERG, and 29 RLQ with powerful outflows. The ratio of beam power to disk luminosity tends to be high for LERG, low for RLQ, and spans the full range of values for HERG. Writing general expressions for the disk luminosity and beam power and applying the empirically determined relationships allows a function that parameterizes the spins of the holes to be estimated. Interestingly, one of the solutions that is consistent with the data has a functional form that is remarkably similar to that expected in the generalized Blandford-Znajek model with a magnetic field that is similar in form to that expected in MAD and ADAF models. Values of the spin function, obtained independent of specific outflow models, suggest that spin and AGN type are not related for these types of sources. The spin function can be used to solve for black hole spin in the context of particular outflow models, and one example is provided. 
\end{abstract}


\section{INTRODUCTION}
A significant amount of theoretical work has shown that the spin of a 
supermassive black hole is likely to impact the 
properties of powerful collimated outflows from supermassive black hole 
systems. For example, the work of Blandford \& Znajek 
(1977), Thorne et al. (1986), Blandford (1990), and 
Meier (1999)  and the more recent work of McKinney \& Gammie (2004), 
McKinney (2005, 2006), Hawley \& Krolik (2006),
Tchekhovskoy et al. (2010), Tchekhovskoy et al. (2011),  
Yuan \& Narayan (2014), and Sadowski et al. (2015) 
to name a few, indicate 
that the spin of the hole may play a significant role in the beam power
of outflows from black hole systems.  

Empirical studies of AGN with measured X-ray iron lines indicate that 
some of the sources have significant spin, as summarized in the 
recent review of Reynolds (2014); selection effects 
may explain this as an observational bias (Reynolds 2015). 
Studies of AGN with outflows analyzed in the 
context of particular models suggest that these sources 
have a broad range of spin values (e.g. Daly 2009, 2011; 
Gnedin et al. 2012; Mikhailov et al. 2015). 

Spin values are likely to indicate whether the accretion history of 
the source under study was chaotic, leading to a low spin
value (King \& Pringle 2006, 2007;  
King et al. 2008),
or more smoothly progressing, leading to a high spin value
(Volonteri et al. 2005; Volonteri et al. 2007; and Dubois
et al. 2014). Thus, 
studies of spins may indicate whether a chaotic
or non-chaotic accretion history was likely to have been dominant for 
that class of source.  

The purpose of the work presented here is to study 
sources with powerful outflows to determine if the
outflow and accretion disk properties of the systems provide 
indications of the spin characteristics of the sources. 
The sample of sources studied is described in section 2,
the analysis and results are presented in section 3, and
a discussion of the results and implications follows in section 4. 

\section{The Sources}
To study the relationship between beam power and accretion disk luminosity,
a sample of sources for which these 
quantities are known was selected. 
Beam powers, or energy per unit time carried by the outflow, 
can be determined from multi-frequency
radio maps of extended powerful FRII (Fanaroff \& Riley 1974) 
(classical double) radio galaxies and 
radio loud quasars, as described by O'Dea et al. 
(2009) and summarized by Daly (2011). The beam powers 
for these sources are not affected
by Doppler beaming and boosting due to bulk motion; 
the sources are large, 
typically much larger than the host galaxy, and 
the radio emission is emitted isotropically. Beam powers are obtained by 
applying the equations of strong shock physics using parameters
empirically determined from the multi-frequency radio maps. 
One parameter is the source age, which is 
determined with a spectral aging analysis. A significant
amount of work has shown that these ages provide reasonable estimates 
for very powerful FRII sources like those included here, 
and uncertainties in the spectral aging analysis are included in 
uncertainties of parameters determined using the analysis, 
as summarized by O'Dea et al. (2009) and references therein, though
there may be some caveats as discussed, for example, by 
Eilek et al. (1997), Blundell \& Rawlings (2000), and 
Hardcastle (2013). 

Thus, the parent population begins with 
classical double radio sources with beam powers that have been 
determined. Sources with beam powers
obtained by O'Dea et al. (2009) and Daly et al. (2012) are included. 
Of these sources, those with a reliable estimate of accretion disk
luminosity were identified; the [OIII] luminosity of the source
was used to determine the bolometric luminosity of the accretion
disk using the well-known relation $L_{bol} = 3500 L_{OIII}$ 
(e.g. Heckman et al. 2004; Dicken et al. 2014). 
Using the tables published by 
Grimes, Rawlings, \& Willott (2004), this led to a sample
of 55 high excitation radio galaxies (HERG), 13 low excitation radio
galaxies (LERG), and 29 radio loud quasars (RLQ). The 
sources types are from Laing, Riley, \& Longair (1983), and the 
black hole masses are from McLure et al. (2004) and McLure et al. (2006).  
\begin{figure}
    \centering
    \includegraphics[width=80mm]{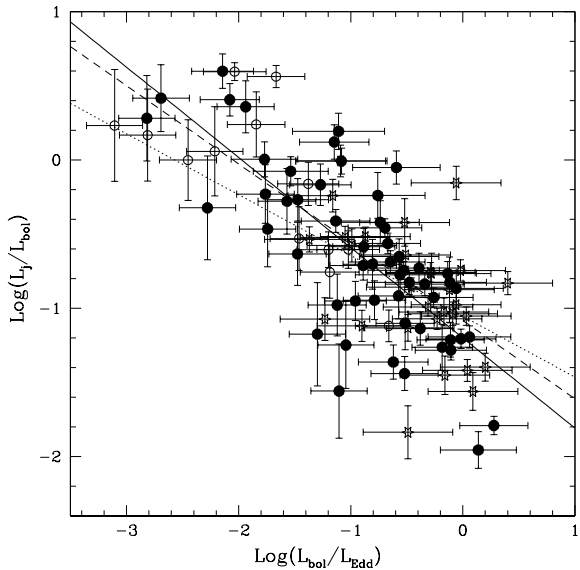}	 
\caption{The Log of the ratio of beam power to accretion disk luminosity 
is shown versus the Log of the ratio of accretion disk luminosity 
to Eddington luminosity for 55 HERG (solid circles), 13 LERG (open circles), 
and 29 RLQ (open stars). 
The best fit lines in (slope, y-intercept) pairs are:
($-0.61 \pm 0.07$, $-1.20 \pm 0.08$) (HERG - solid line); 
($-0.53 \pm 0.15$, $-1.08 \pm 0.29$) (LERG - dashed line);  
($-0.41 \pm 0.15$, $-1.05 \pm 0.10$) (RLQ - dotted line); and
($-0.56 \pm 0.05$, $-1.14 \pm 0.06$) (all sources). All fits are unweighted and 
the same symbols are used in all figures.}
		  \label{fig:F1}
    \end{figure} 
\section{Analysis}
The ratio of the beam power to the accretion disk luminosity is a
fundamental physical variable that parameterizes the strength
of the outflow relative to the accretion disk. This ratio is shown 
in Fig. 1 as
a function of the accretion disk luminosity 
normalized by the Eddington luminosity, $L_{Edd} \simeq 1.3 \times
10^{46} M_8 \hbox{ erg s}^{-1}$ where $M_8$ is the black hole mass in units of
$10^8 M_{\odot}$. 
There is an obvious trend between these two quantities, and the best
fit parameters are listed in the figure caption. To 
test whether this is a spurious result due to the Malmquist bias
(e.g. Feigelson \& Berg 1983) a partial correlation analysis was
carried out using the code of Akritas \& Siebert (1996). 
Following Hardcastle et al. (2009) and Mingo et al. (2014), 
redshift is used as a proxy 
for distance and a ratio of partial Kendall's $\tau$ 
to the square root of the variance $\sigma$ of $\tau/ \sigma >3$ indicates 
a significant correlation between $\rm{Log}(L_j/L_{bol})$ and 
$\rm{Log}(L_{bol}/L_{Edd})$ 
in the presence of $\rm{Log}(1+z)$. The ratio of $\tau/ \sigma$ is 
$12.6$ for all sources; 
$10.5$ for HERG; $1.9$ for LERG; and $3.2$ for RLQ. 
This indicates the fits obtained are valid with the possible exception
of that obtained for LERG sources. 

These fits suggest that  
\begin{equation}
{L_j \over L_{bol}} \propto \left({L_{bol} \over L_{Edd}}\right)^{\alpha_*}~ 
\end{equation}
with $\alpha_* = -0.5$ 
provides a good description of the data, with the possible exception
of the LERG data. The impact of using 
$\alpha_* =  -0.56 \pm 0.05$ is discussed below. 

It is convenient to parameterize the accretion disk luminosity as 
$L_{bol} \propto \epsilon ~ \dot{M} \propto \epsilon ~\dot{m} ~M$, where 
$M$ is the black hole mass, 
$\dot{M}$ is the mass accretion rate, $\dot{m} \equiv \dot{M}/\dot{M}_{Edd}$
is the dimensionless mass accretion rate, 
$\dot{M}_{Edd} \equiv L_{Edd} c^{-2}$ is the Eddington accretion rate, and 
$\epsilon$ is a dimensionless efficiency factor.
It is convenient to parameterize the beam power as
$L_j \propto \dot{m}^a ~M^b~f(j)$, where $f(j)$ is a function 
of the spin of the black hole. To determine the values of $a$ and $b$,
empirical relationships are considered. Eq. (1) with $\alpha_* = -0.5$ indicates
that $\dot{m}^a ~M^b~f(j) \propto (\epsilon ~\dot{m})^{1/2} M$, which suggests
that $b=1$. In this case, the ratio of $L_j/L_{bol}$ is expected to be
independent of black hole mass. As illustrated in Fig. 2, 
the data are consistent with the ratio being independent
of mass, so we adopt a value of $b=1$. Thus,    
$\dot{m}^a ~M~f(j) \propto (\epsilon ~\dot{m})^{1/2} M$, 
which indicates that $\dot{m}^a \propto (\epsilon ~\dot{m})^{1/2}$.  
The two simplest solutions to this equation are $a=1$ with 
$\epsilon \propto \dot{m}$ and $a=1/2$ with $\epsilon$ = constant. 
The first solution yields $L_j \propto \dot{m} ~M~f(j)$
with $L_{bol} \propto \dot{m}^2 ~M$. The second solution yields 
$L_j \propto \dot{m}^{1/2} ~M~f(j)$ with $L_{bol} \propto \dot{m} ~M$. 
It is interesting to note that the expression for $L_j$ indicated
by the first solution is very similar to that expected in some  
models of jet production, as discussed in section 4. 

\begin{figure}
    \centering
    \includegraphics[width=80mm]{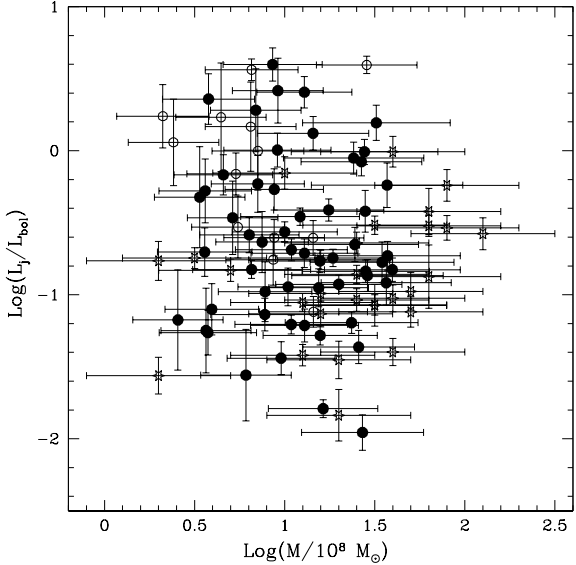}			 
\caption{The Log of the ratio of beam power to accretion disk luminosity
is shown versus the Log of black hole mass. There is no correlation between
these quantities.  The slopes of the best fit lines are:
$-0.07 \pm 0.24$ (HERG); 
$-0.40 \pm 0.51$ (LERG);
$0.26 \pm 0.17$ (RLQ);  and
$-0.17 \pm 0.14$ (all sources).}		  
\label{fig:F2}
    \end{figure} 

The general equations for $L_{bol}$ and $L_j$ can be combined to 
solve for the function $f(j)$. To do this requires that constants of 
proportionality be obtained for $L_{bol}$ and $L_j$. This may be
done by parameterizing the maximum possible emission in 
terms of the Eddington luminosity, so the maximum
possible value of $L_{bol}$ is 
$L_{bol}(max) = g_{bol} L_{Edd}$ and that for $L_j$ is 
$L_j(max) = g_j L_{Edd}$. Assuming the maximum values
are reached when $\dot{m} = 1$ and $\epsilon = 1$, and absorbing
all constants of proportionality into the coefficients yields 
\begin{equation}
L_{bol,44} \simeq ~130 ~g_{bol}~ (\epsilon \dot{m}) M_8
\end{equation}
\begin{equation} 
L_{j,44} \simeq ~130 ~g_j~ \dot{m}^a M_8f(j)/f_{max}.
\end{equation}
Here $L_{bol,44}$ and $L_{j,44}$ are
in units of $10^{44} \hbox{ erg s}^{-1}$ and 
$f_{max}$ is the maximum possible value of the function $f(j)$,
which is typically obtained when the dimensionless spin $j = 1$;
here $j$ (sometimes denoted $a$ or $a_*$) 
is defined in the usual way, $j \equiv Jc/(GM^2)$, where $J$ is the 
spin angular momentum of the hole. Combining eqs. (2) and (3), and  
using the relationship
$\dot{m}^a = (\epsilon \dot{m})^{1/2}$ indicated above, 
we obtain  
\begin{equation}
{f(j) \over f_{max}} \simeq \left({L_{j,44} \over g_j}\right)~\left({g_{bol} 
\over 130 L_{bol,44} M_8}\right)^{1/2}~,
\end{equation}
independent of the value of $a$, and thus independent of specific outflow 
models. 

\begin{figure}
    \centering
    \includegraphics[width=80mm]{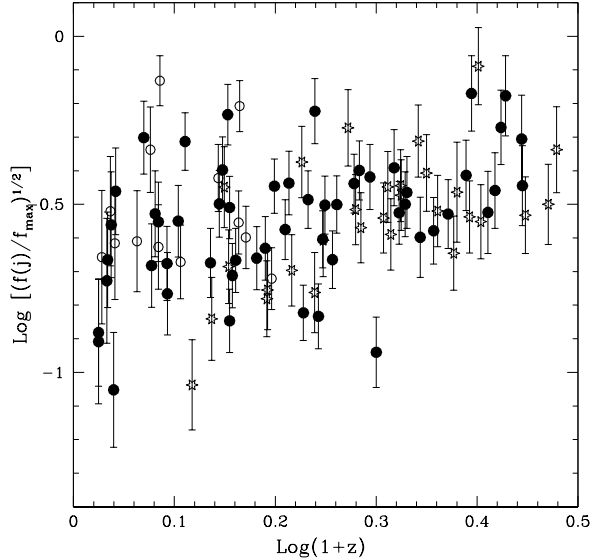}
\caption{The Log of the square root of the spin function is shown versus 
Log of (1+z);
this is expected to be a reasonable first order
approximation of spin. The values shown here are obtained
using eq. (4) with $g_{bol} = 1$ and $g_j=1$. }
		 \label{fig:F3}
    \end{figure}

If this analysis is carried out for the specific value of $\alpha_*$ 
obtained for all sources, $\alpha_* = -0.56 \pm 0.05$, 
then eq. (4) becomes
\begin{equation}
{f(j) \over f_{max}} \simeq \left({L_{j,44} \over g_j}\right)
\left({g_{bol} \over L_{bol,44}}\right)^{0.44 \pm 0.05}(130 M_8)^{-0.56 \pm 0.05}~, 
\end{equation}
which is also independent of $a$, and is very similar to eq. (4). 

Empirical results for $\sqrt{f(j)/f_{max}}$ obtained using eq. (4) 
with $g_{bol} = 1$ and $g_j = 1$ are shown in Fig. 3 and listed in Table 1. 
The square root of the function is shown because in many models this
is a good first order approximation to the spin of the hole. 
Estimates of black hole spins can be 
obtained from  $\sqrt{f(j)/f_{max}}$ in the context of 
specific models. 

Changing the normalizations $g_{bol}$ and $g_j$ will cause the values
of $f(j)/f_{max}$ to shift. However, only small upward shifts are allowed by 
the data. This follows because, empirically  
(e.g. see Fig. 4), $g_{bol}$ must be close to one, and
certainly can not be much less than one, and $f(j) \propto g_{bol}^{1/2}$. 
And, as the value of 
$g_j$ decreases, $f(j)/f_{max}$ increases. Requiring that 
the largest values of 
$f(j)/f_{max}$ remain less than or equal to one indicates that 
$g_j$ should be greater than about 0.4, consistent
with the empirical results illustrated in Fig. 4.
Thus, the true value
of $f(j)/f_{max}$ is only allowed to float between the value obtained
here, and the current value divided by about 0.4.

If a particular outflow model is specified, the value of $j$ may
be obtained. For example, in one representation of the generalized
Blandford-Znajek (BZ) model
$\sqrt{f(j)/f_{max}} = j (1+\sqrt{1-j^2})^{-1}$ 
(e.g. Blandford \& Znajek 1977; Tchekhovskoy et al. 2010; 
Yuan \& Narayan 2014). 
Values of $j$ obtained in this model are shown Fig. 5 for 
$g_{bol} = 1$ and $g_j = 1$. 

\begin{figure}
    \centering
    \includegraphics[width=80mm]{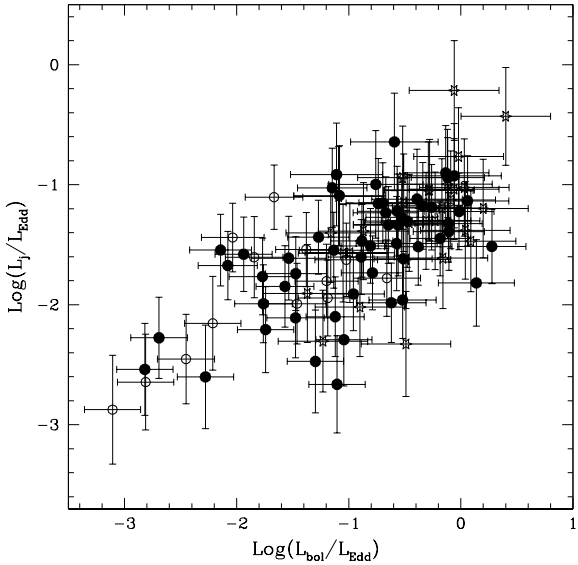}	
\caption{The Log of the beam power (in Eddington units) is  
shown versus the Log of the accretion disk luminosity (in Eddington units)
so that the range of values can be seen.}
	  \label{fig:F4}
    \end{figure}

\section{Discussion}
There is a clear separation of sources
in terms of the ratio of beam power to disk luminosity 
(see Fig. 1). LERG sources tend to have the highest ratio of
beam power to disk luminosity, with some sources having a
beam power comparable to or even larger than the disk luminosity.
RLQ tend to have a low ratio, with most sources having a beam power
less than about 10 \% the disk luminosity. HERG tend to span
the full range of values of this ratio. The relationship between
this ratio and the Eddington normalized disk luminosity is 
statistically significant after accounting for the dependence of
these quantities on redshift for the HERG, RLQ, all sources combined,
but not for the LERG sources.  

\begin{figure}
    \centering
    \includegraphics[width=80mm]{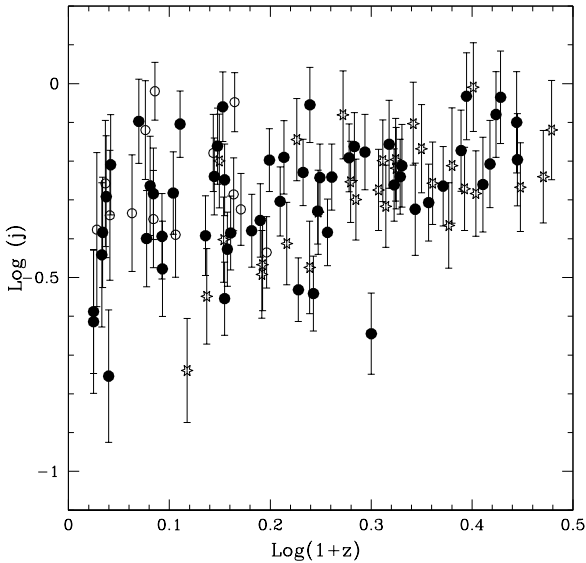}
\caption{The Log of the black hole spin, obtained 
in the context of the generalized BZ model, is shown
versus the Log of (1+z).}
		  \label{fig:F5}
    \end{figure}

Solutions that are consistent with the data 
may be compared with 
theoretical expectations. In the generalized BZ model, the
equation for beam power is $L_j = \kappa_j B^2 M^2 f(j)$ 
(e.g. Blandford \& Znajek 1977; Blandford 1990; 
Tchekhovskoy et al. 2010). 
For ADAF and MAD accretion disks,  
the magnetic field strength depends on multiple parameters
including $\dot{m}$ and $M$; the dependence of the field
on these parameters is $B^2 \propto (\dot{m}/M)$ (e.g. Yuan \& Narayan 2014)
or $B_4^2 = g^2_B (\dot{m}/M_8)$, 
indicating that $L_j \propto \dot{m} ~M~f(j)$;
$B_4$ is the field strength in units of $10^4$ G. 
Interestingly, this theoretical
equation for the beam power is identical to the empirical 
relation indicated by the first solution discussed in section 3. 
Thus, the generalized BZ model with an ADAF or MAD accretion disk
is consistent with the empirically motivated relationships between
the beam power, disk luminosity, and Eddington luminosity 
discussed in section 3, perhaps suggesting that this is in
fact the correct model, or close to the correct model. In this context, 
the first solution along with the normalization given by 
equation (3) allows a determination of the coefficient 
$g_B$ that describes the field strength. 
Obtaining the normalization $\kappa_j$ from Tchekhovskoy et al. (2011), which 
is nearly identical to that obtained by Daly (2009) 
with $j^2$ replaced by $f(j)$, we obtain $g_B \simeq 20 \sqrt{g_j}$. This  
is similar to the value expected in the MAD model
of about $g_B \simeq 30$ (e.g. Yuan \& Narayan 2014)
accounting for their different definition of $\dot{M}_{EDD}$. 
Note that to be consistent with this representation of
$L_j$, we require $L_{bol} \propto \dot{m}^2 M_8$. Theoretical
representations of $L_j$ and $L_{bol}$ should be consistent
with the empirical relation indicated by eq. (1).  

The spin function $f(j)/f_{max}$ may be obtained independent
of specific outflow models; this quantity is expected to provide 
a good first order estimate of black hole spin.
A broad range of values of $\sqrt{f(j)/f_{max}}$ is obtained (see Fig. 3). 
The values and range of values of $\sqrt{f(j)/f_{max}}$ are similar for 
all three types of sources studied; sources do not separate 
out according to this quantity. This suggests that spin is not
related to AGN type for FRII HERG and RLQ, and possibly also for LERG.

Finally, even though  $\sqrt{f(j)/f_{max}}$ depends
upon the values of $g_{bol}$ and $g_j$, it is argued in section 3
that only small upward shifts of this quantity are allowed by the 
data. At this point, the data are not sufficient to be able to distinguish
between chaotic accretion and non-chaotic 
accretion for the sources studied; both
are consistent with the results obtained here. Further studies of this
type with larger numbers of sources may be able to distinguish between
these accretion scenarios. 

\begin{table*}
\begin{minipage}{140mm}
\scriptsize
\caption{Black Hole Spin Function of FRII Sources}   
\label{tab:comp}        
\begin{tabular}{lllrlllr}   
\hline\hline                    

Source &  type& z & Log$(\sqrt{f(j)/f_{max}})$ & Source &  type& z & Log$(\sqrt{f(j)/f_{max}})$\\
\hline                          
3C	33	&	HERG	&	0.059	&$	-0.88	\pm	0.16	$	&	3C	322	&	HERG	&	1.681	&$	-0.18	\pm	0.12	$	\\
3C	192	&	HERG	&	0.059	&$	-0.91	\pm	0.19	$	&	3C	239	&	HERG	&	1.781	&$	-0.31	\pm	0.13	$	\\
3C	285	&	HERG	&	0.079	&$	-0.73	\pm	0.19	$	&	3C	294	&	HERG	&	1.786	&$	-0.45	\pm	0.12	$	\\
3C	452	&	HERG	&	0.081	&$	-0.67	\pm	0.14	$	&	3C	225B	&	HERG	&	0.582	&$	-0.45	\pm	0.08	$	\\
3C	388	&	HERG	&	0.090	&$	-0.56	\pm	0.16	$	&	3C	55	&	HERG	&	0.735	&$	-0.23	\pm	0.10	$	\\
3C	321	&	HERG	&	0.096	&$	-1.06	\pm	0.17	$	&	3C	68.2	&	HERG	&	1.575	&$	-0.53	\pm	0.12	$	\\
3C	433	&	HERG	&	0.101	&$	-0.46	\pm	0.13	$	&	3C	35	&	LERG	&	0.067	&$	-0.66	\pm	0.20	$	\\
3C	20	&	HERG	&	0.174	&$	-0.31	\pm	0.11	$	&	3C	326	&	LERG	&	0.088	&$	-0.52	\pm	0.16	$	\\
3C	28	&	HERG	&	0.195	&$	-0.69	\pm	0.12	$	&	3C	236	&	LERG	&	0.099	&$	-0.62	\pm	0.17	$	\\
3C	349	&	HERG	&	0.205	&$	-0.53	\pm	0.13	$	&	4C	12.03	&	LERG	&	0.156	&$	-0.61	\pm	0.15	$	\\
3C	436	&	HERG	&	0.214	&$	-0.56	\pm	0.12	$	&	3C	319	&	LERG	&	0.192	&$	-0.34	\pm	0.13	$	\\
3C	171	&	HERG	&	0.238	&$	-0.68	\pm	0.11	$	&	3C	132	&	LERG	&	0.214	&$	-0.63	\pm	0.13	$	\\
3C	284	&	HERG	&	0.239	&$	-0.77	\pm	0.12	$	&	3C	123	&	LERG	&	0.218	&$	-0.14	\pm	0.07	$	\\
3C	300	&	HERG	&	0.270	&$	-0.55	\pm	0.11	$	&	3C	153	&	LERG	&	0.277	&$	-0.67	\pm	0.11	$	\\
3C	438	&	HERG	&	0.290	&$	-0.32	\pm	0.09	$	&	4C	14.27	&	LERG	&	0.392	&$	-0.43	\pm	0.10	$	\\
3C	299	&	HERG	&	0.367	&$	-0.68	\pm	0.10	$	&	3C	200	&	LERG	&	0.458	&$	-0.56	\pm	0.09	$	\\
3C	42	&	HERG	&	0.395	&$	-0.50	\pm	0.10	$	&	3C	295	&	LERG	&	0.461	&$	-0.21	\pm	0.08	$	\\
3C	16	&	HERG	&	0.405	&$	-0.40	\pm	0.10	$	&	3C	19	&	LERG	&	0.482	&$	-0.60	\pm	0.09	$	\\
3C	274.1	&	HERG	&	0.422	&$	-0.24	\pm	0.09	$	&	3C	427.1	&	LERG	&	0.572	&$	-0.72	\pm	0.09	$	\\
3C	244.1	&	HERG	&	0.428	&$	-0.85	\pm	0.09	$	&	3C	249.1	&	RLQ	&	0.311	&$	-1.04	\pm	0.13	$	\\
3C	457	&	HERG	&	0.428	&$	-0.51	\pm	0.09	$	&	3C	351	&	RLQ	&	0.371	&$	-0.84	\pm	0.12	$	\\
3C	46	&	HERG	&	0.437	&$	-0.72	\pm	0.10	$	&	3C	215	&	RLQ	&	0.411	&$	-0.45	\pm	0.12	$	\\
3C	341	&	HERG	&	0.448	&$	-0.67	\pm	0.10	$	&	3C	47	&	RLQ	&	0.425	&$	-0.69	\pm	0.11	$	\\
3C	172	&	HERG	&	0.519	&$	-0.66	\pm	0.09	$	&	3C	334	&	RLQ	&	0.555	&$	-0.78	\pm	0.11	$	\\
3C	330	&	HERG	&	0.549	&$	-0.63	\pm	0.09	$	&	3C	275.1	&	RLQ	&	0.557	&$	-0.76	\pm	0.12	$	\\
3C	49	&	HERG	&	0.621	&$	-0.58	\pm	0.09	$	&	3C	263	&	RLQ	&	0.646	&$	-0.70	\pm	0.11	$	\\
3C	337	&	HERG	&	0.635	&$	-0.44	\pm	0.09	$	&	3C	207	&	RLQ	&	0.684	&$	-0.38	\pm	0.11	$	\\
3C	34	&	HERG	&	0.690	&$	-0.83	\pm	0.08	$	&	3C	254	&	RLQ	&	0.734	&$	-0.77	\pm	0.12	$	\\
3C	441	&	HERG	&	0.708	&$	-0.49	\pm	0.09	$	&	3C	175	&	RLQ	&	0.768	&$	-0.61	\pm	0.11	$	\\
3C	247	&	HERG	&	0.749	&$	-0.84	\pm	0.10	$	&	3C	196	&	RLQ	&	0.871	&$	-0.28	\pm	0.11	$	\\
3C	277.2	&	HERG	&	0.766	&$	-0.61	\pm	0.08	$	&	3C	309.1	&	RLQ	&	0.904	&$	-0.52	\pm	0.10	$	\\
3C	340	&	HERG	&	0.775	&$	-0.51	\pm	0.09	$	&	3C	336	&	RLQ	&	0.927	&$	-0.57	\pm	0.10	$	\\
3C	352	&	HERG	&	0.806	&$	-0.67	\pm	0.09	$	&	3C	245	&	RLQ	&	1.029	&$	-0.54	\pm	0.10	$	\\
3C	263.1	&	HERG	&	0.824	&$	-0.50	\pm	0.09	$	&	3C	212	&	RLQ	&	1.049	&$	-0.45	\pm	0.11	$	\\
3C	217	&	HERG	&	0.897	&$	-0.44	\pm	0.09	$	&	3C	186	&	RLQ	&	1.063	&$	-0.59	\pm	0.11	$	\\
3C	175.1	&	HERG	&	0.920	&$	-0.40	\pm	0.09	$	&	3C	208	&	RLQ	&	1.110	&$	-0.45	\pm	0.11	$	\\
3C	289	&	HERG	&	0.967	&$	-0.42	\pm	0.10	$	&	3C	204	&	RLQ	&	1.112	&$	-0.48	\pm	0.10	$	\\
3C	280	&	HERG	&	0.996	&$	-0.94	\pm	0.10	$	&	3C	190	&	RLQ	&	1.197	&$	-0.32	\pm	0.11	$	\\
3C	356	&	HERG	&	1.079	&$	-0.39	\pm	0.11	$	&	3C	68.1	&	RLQ	&	1.238	&$	-0.41	\pm	0.11	$	\\
3C	252	&	HERG	&	1.103	&$	-0.53	\pm	0.09	$	&	4C	16.49	&	RLQ	&	1.296	&$	-0.52	\pm	0.11	$	\\
3C	368	&	HERG	&	1.132	&$	-0.50	\pm	0.10	$	&	3C	181	&	RLQ	&	1.382	&$	-0.65	\pm	0.11	$	\\
3C	267	&	HERG	&	1.140	&$	-0.47	\pm	0.11	$	&	3C	268.4	&	RLQ	&	1.400	&$	-0.47	\pm	0.15	$	\\
3C	324	&	HERG	&	1.206	&$	-0.60	\pm	0.12	$	&	3C	14	&	RLQ	&	1.469	&$	-0.54	\pm	0.11	$	\\
3C	266	&	HERG	&	1.275	&$	-0.58	\pm	0.10	$	&	3C	270.1	&	RLQ	&	1.519	&$	-0.09	\pm	0.11	$	\\
3C	13	&	HERG	&	1.351	&$	-0.53	\pm	0.10	$	&	3C	205	&	RLQ	&	1.534	&$	-0.55	\pm	0.11	$	\\
4C	13.66	&	HERG	&	1.450	&$	-0.42	\pm	0.10	$	&	3C	432	&	RLQ	&	1.805	&$	-0.54	\pm	0.11	$	\\
3C	437	&	HERG	&	1.480	&$	-0.17	\pm	0.11	$	&	3C	191	&	RLQ	&	1.956	&$	-0.50	\pm	0.12	$	\\
3C	241	&	HERG	&	1.617	&$	-0.46	\pm	0.11	$	&	3C	9	&	RLQ	&	2.012	&$	-0.34	\pm	0.13	$	\\
3C	470	&	HERG	&	1.653	&$	-0.27	\pm	0.11	$	\\												
														
\end{tabular}
\end{minipage}
\end{table*}

\section*{Acknowledgments}
It is a pleasure to thank 
the organizers of conference 
``The Physics of Supermassive Black Hole Formation and Feedback'' 
where this work was presented; 
Sasha Tchekovskoy and Johnathan McKinney for 
interesting and helpful discussions;  
and the referee for helpful comments
and suggestions. 
This work was supported in part by Penn State University.

\end{document}